\documentclass[twocolumn]{maeb2015}
\usepackage[spanish]{babel}
\usepackage[latin1]{inputenc}
\usepackage{graphicx}

\usepackage[onelanguage,ruled,norelsize,boxed]{algorithm2e}
\usepackage{graphicx}
\usepackage{graphics}
\usepackage{subfig}
\usepackage{multirow}
\usepackage{epstopdf}
\usepackage{bigstrut}

\def\BibTeX{{\rm B\kern-.05em{\sc i\kern-.025em b}\kern-.08em
    T\kern-.1667em\lower.7ex\hbox{E}\kern-.125emX}}

\begin{document}

\title{BVNS para el problema del \\bosque generador \emph{k}-etiquetado}

\author{Sergio Consoli$^{(1)}$,\thanks{$^{(1)}$
National Research Council (CNR),
ISTC/STLab, 95126 Catania (CT), Italia
E-mail: sergio.consoli@istc.cnr.it}
Nenad Mladenovi\'c$^{(2)}$\thanks{$^{(2)}$
School of Information Systems, Computing and Mathematics,
Brunel University, London, UK
E-mail: nenad.mladenovic@brunel.ac.uk}
José A. Moreno-Pérez$^{(3)}$\thanks{$^{(3)}$
Departamento de Ingeniería Informática, IUDR,
Universidad de La Laguna,
38271 La Laguna, España
E-mail: jamoreno@ull.es}}

\maketitle

\begin{abstract}

En este trabajo proponemos una VNS eficiente para el problema del bosque generador de $k$-etiquetado.
Este problema es una extensión del problema del árbol generador de mínimo etiquetado ({\sl Minimum Labelling Spanning Tree Problem})
con importantes aplicaciones en redes de telecomunicaciones y transporte multimodal.
Se trata de, dado un grafo no dirigido cuyos enlaces están etiquetados y un número entero positivo ${k}$
encontrar el bosque o subgrafo con el menor número de componentes conexas usando a lo sumo $k$ etiquetas distintas.
Para abordar lo propone una VNS básica (BVNS, \emph{Basic Variable Neighbourhood Search}) donde la amplitud máxima del proceso de agitación es el parámetro clave.
Se estudian diferentes estrategias para establecer el valor de dicho parámetro obteniéndose como mejor estrategia hacerlo depender del estado del proceso de búsqueda.
La BVNS con la estrategia seleccionada es experimentalmente comparada con las metaheurísticas que han aparecido en la literatura aplicada a este tipo de problema.

\end{abstract}


\section{Introducción}
\label{Introduction}

Las redes de transporte multimodal pueden representarse por grafos donde cada enlace tiene una etiqueta que denota una empresa que lo gestiona o tipo de transporte diferente.
En  muchos contextos de transporte multimodal se pretende garantizar la mayor conectividad posible entre los nodos de la red utilizando para ello el número mínimo de empresas proveedoras o tipos de transporte \cite{Miller2006}, \cite{Vannes}.
El propósito es reducir el coste de la implantación del sistema de transporte o la complejidad global de la red.
En algunas situaciones reales estaremos interesados en la optimización de la conectividad con un límite en el número de proveedores o modos de transporte utilizados \cite{Miller2006}.
Este problema denominado problema del bosque generador etiquetado ({\emph Labelled Spanning Forest Problem}, LSFP), propuesto originalmente en ~\cite{CerulliKLSF},
pertenece a una clase de problemas derivados del problema del árbol generador de etiquetado mínimo
(\emph{ Minimum Labelling Spanning Tree Problem}, MLSTP)~\cite{Chang}, que ha recibido bastante atención recientemente.
El problema del bosque generador etiquetado, con un entero positivo ${k}$ como cota superior del número de etiquetas a utilizar, se denomina problema del bosque generador $k$-etiquetado.
Se denota por $k$LSFP ($k$-{\emph Labelled Spanning Forest Problem}) y se formaliza en los siguientes términos:

Sea ${k}$ un entero positivo y $G=(V,A,L)$ un grafo no dirigido etiquetado donde $V$ es el conjunto de sus vértices y
$A$ el de sus aristas que están etiquetadas con etiquetas del conjunto finito de etiquetas $L$.
Se trata de encontrar el conjunto de etiquetas $L^* \subseteq L$ de no más de ${k}$ etiquetas distintas tal que el
subgrafo $G^*$ de $G$ tomando todas las aristas cuya etiquetas están en $L^*$ tenga el menor número posible de componentes conexas.
Formalmente, denotamos por $l(a) \in L$ la etiqueta de la arista $a\in A$.
Para cualquier conjunto de etiquetas $C \subseteq L$ denotemos por $G(C)$ al subgrafo del grafo etiquetado $G$ formado por todas las aristas con etiquetas en $C$;
es decir $G(C) = (V,A(C))$ donde $A(C) = \{ a\in A: l(a)\in C\}$.
Denotemos por $Comp(C)$ el número de componentes conexas del subgrafo $G(C)$. El problema es:
$$\min \{ Comp(C) : C \subseteq L, |C| \le {k}\} .$$
Como solución al problema se aporta, junto al conjunto de etiquetas óptimo $L^*$, el grafo asociado $G^* = G(V,A(L^*))$ o un bosque generador $H^*$;
es decir un subgrafo con el mismo número de componentes conexas que $G^*$ pero sin ciclos.
Si la solución óptima $G^*$ contiene ciclos se pueden romper arbitrariamente cada uno de ellos quitando una arista cualquiera en tiempo polinomial para obtener el bosque generador $H^*$.

En ~\cite{CerulliKLSF} se prueba que el problema del bosque generador $k$-etiquetado, $k$LSFP, es un problema $NP$-duro.
El $k$LSFP generaliza el problema del árbol generador de mínimo etiquetado (MLSTP, {\sl Minimum Labelling Spanning Tree Problem}) cuyo objetivo es determinar el árbol generador del grafo no dirigido etiquetado con el número mínimo de etiquetas. Una solución al problema MLST sería también una solución al problema $k$LSF si el árbol solución obtenido no sobrepasa el límite de ${k}$ en el número de etiquetas utilizadas.
Por lo tanto los métodos de solución del problema $k$LSF se deben buscar en primer lugar adaptando los propuestos para el MLSTP.
En ~\cite{Chang} se propuso el algoritmo MVCA (\emph{ Maximum Vertex Covering Algorithm}) como primera heurística para el problema MLST que fue mejorada por \cite{Krumke};
otras propuestas metaheurísticas para dicho problema aparecen en ~\cite{Golden3,Cerulli1,Cerulli2,MLSTP-EJOR,MLSteiner-ANOR,Golden2,Chwatal2010,MLSTP-hybrid,bruggemann}.

La estructura del resto del trabajo es como sigue.
En la Sección~\ref{otrosalg} se presentan los detalles de los métodos de solución para el $k$LSFP aparecidos en la literatura propuestos por diversos autores.
En la Sección~\ref{sec-vns} se describe con algo más de detalle la VNS básica propuesta y el ajuste de la amplitud de agitación para el problema.
La Sección~\ref{resultados} muestra algunos resultados de la comparación experimental y finalizamos
con la sección de conclusiones~\ref{sec-conclusions}.

\section{Algoritmos para el $k$LSFP} \label{otrosalg}

Se describe, en primer lugar, un método exacto derivado de ~\cite{MLSTP-EJOR} para el $kLSF$. A continuación se describen las principales metaheurísticas aparecidas en la literatura:
un Método Piloto ~\cite{Cerulli2,CerulliKLSF} y un algoritmo genético ~\cite{Golden3,CerulliKLSF}.
Se presenta la adaptación al $k$LSFP del procedimiento de búsqueda voraz aleatorizada adaptativa GRASP propuesta en ~\cite{MLSTP-EJOR} para el MLSTP.

Todos estos algoritmos se basan, de una manera u otra, en el algoritmo MVCA ~\cite{Chang,Krumke} adaptado al $k$LSF por ~\cite{CerulliKLSF} como subprocedimiento constructivo.
A partir de un conjunto inicialmente vacío de etiquetas $C$, el MVCA añade iterativamente a $C$ la etiqueta $l \in (L - C)$ tal que el subgrafo $G(C) = (V, A(C))$
tenga el número mínimo de componentes conexas $|Comp(C)|$, y se detiene cuando el número de etiquetas utilizadas alcance la cota superior ($|C| = k$), o cuando quede sólo una componente conexa ($|Comp(C)| = 1$).

\subsection{Método Exacto} \label{subsec-exact}

El enfoque exacto para el $k$LSF se basa en un proceso de exploración por vuelta atrás (\emph{backtracking}).
Dado el grafo no dirigido etiquetado $ G = (V, A, L) $ con $ n $ vértices, $ m $ aristas y  $ \ell $ etiquetas,
y un número entero positivo  ${k} $ como límite superior para el número de etiquetas,
el método exacto realiza una ramificación y poda
en el espacio de soluciones parciales basada en un procedimiento recursivo.
Se parte de un conjunto de etiquetas vacío y se construye una solución mediante la adición iterativa de etiquetas hasta que el número de etiquetas llegue a $ {k} $.
Cuando se alcanza la cota se aplica la fase de vuelta atrás para explorar nuevas alternativas.
El algoritmo se detiene si se encuentra una solución con sólo una componente conexa sin superar el número $k$ de etiquetas utilizadas.
Este proceso da lugar a una enumeración de todas las posibles combinaciones de $ {k} $ etiquetas.
Por lo tanto, la combinación con el menor número de componentes conexas es la solución exacta proporcionada como resultado del método.
Aunque el tiempo de cómputo de este procedimiento es exponencial en el número de etiquetas, si el tamaño del problema es moderado
o el valor de la solución óptima es pequeño, el tiempo de ejecución de este método exacto es razonable y es posible obtener la solución exacta.

\subsection{Método Piloto} \label{subsec-pilot}

La idea central del Método Piloto (PM)~\cite{Cerulli2,CerulliKLSF}
consiste en agotar todas las opciones posibles con respecto a una llamada solución maestra, por medio de una heurística básica.
Esta heurística básica realiza tentativamente iteraciones con respecto a la solución maestra hasta que se evalúan todas las posibles elecciones locales.
La mejor solución hasta entonces se convierte en la nueva solución maestra, y el procedimiento continúa hasta que se cumplan las condiciones de parada predefinidas,
como alcanzar cierto tiempo máximo permitido de CPU o cierto número máximo de iteraciones.

El Método Piloto para el problema $k$LSF parte de la solución nula (un conjunto vacío de etiquetas) como solución maestra $ M $.
Entonces, para cada etiqueta $ i \notin M $, se trata tentativamente de extender una copia de $ M $ a una solución completa
incluyendo la etiqueta $i$, construida mediante el uso del algoritmo MVCA como heurística básica.
Se inserta en la solución parcial la etiqueta que más disminuye el número componentes conexas del subgrafo parcial con una estrategia voraz (\emph{greedy}) ~\cite{CerulliKLSF}.
El número de componentes conexas de la solución obtenida a partir de $ M \leftarrow M\cup \{i \} $ se usa como función objetivo de cada
candidato $ i \notin M $. Cuando todas las etiquetas candidatas posibles con respecto a la solución maestra han sido evaluadas, la etiqueta candidata
$ i^*$ con menor valor de la función objetivo se añade a la solución maestra ($ M \leftarrow M \cup \{i^* \} $).
Sobre la base de esta nueva solución maestra $ M $, se inician nuevas iteraciones
$ \forall i \notin M $, proporcionando un nuevo elemento de la solución de $ i^* $, y así sucesivamente.
Este mecanismo se itera hasta que no se necesite añadir más etiquetas a la solución maestra
(es decir, hasta que el número de etiquetas utilizadas $ | M | $ alcanza $ {k} $, o se obtenga una única componente conexa).
Alternativamente, se pueden imponer algunas condiciones de finalización que permitan que el algoritmo pueda alcanzar estas condiciones.
La última solución maestra corresponde a la mejor solución hasta entonces y se aporta como respuesta del método.

\subsection{Algoritmo Genético} \label{mga-sec}

Los Algoritmos Genéticos (AGs) se inspiran en los principios de la selección natural, la adaptación al entorno, la evolución y
operaciones de cruce y mutación entre los individuos dentro de una población ~\cite{GAs}.
Para la aplicación de esta metaheurística al $k$LSFP, dado el grafo etiquetado no dirigido $ G = (V, A, L) $, un individuo (o cromosoma) de la población
es un subgrafo de $G$, donde cada posible etiqueta se corresponde un gen, y el grado de adaptación (\emph{fitness}) se define como el número de componentes conexas ~\cite{Golden3,CerulliKLSF}.
La población inicial se genera añadiendo etiquetas al azar a un conjunto vacío, hasta que aparecen individuos con $ {k}$ etiquetas.
Los operadores de cruce y mutación se aplican a continuación para dar lugar a una nueva generación desde la anterior ~\cite{Golden3,CerulliKLSF}.
Las probabilidades de cruce y mutación se establecen al 100\%~\cite{Golden3,CerulliKLSF}.
El número de generaciones se establece como la mitad del tamaño de la población.
Por lo tanto, en este algoritmo genético ~\cite{Golden3,CerulliKLSF} el único parámetro a ajustar es el número de generaciones o el tamaño de la población.

El cruzamiento da lugar a un único descendiente para cada dos soluciones factibles padres.
Dados los dos padres $ P_1 \subset L $ y $ P_2 \subset L $, se comienza formando su unión $ P = P_1 \cup P_2 $.
A continuación se agregan etiquetas de $ P $ a una solución inicialmente vacía
utilizando la estrategia del MVCA hasta que se obtenga una solución hija con $ {k} $ etiquetas.
Es decir, se aplica el MVCA al subgrafo $G(P)$ de etiquetas en $P$.
Por otro lado, la operación de mutación
consiste en una perturbación: se añade una nueva etiqueta a al azar,
y se elimina la etiqueta que dé lugar al individuo con mejor ajuste.

Tras de una serie de generaciones el algoritmo converge y el mejor individuo de la población se proporciona como resultado.
Téngase en cuenta que el algoritmo genético termina antes de la última generación si se encuentra una solución factible formada por una sola componente conexa.

\subsection{Búsqueda Adaptativa Voraz Aletaorizada} \label{subsec-grasp}

El procedimiento de búsqueda adaptativa voraz aletaorizada conocida como GRASP ({\sl Greedy Randomized Adaptiva Search Procedure})
es básicamente una metaheurística constructiva que consta de una fase de iterativa constructiva seguida de una fase de mejora basada en una búsqueda local.
En la fase de constructiva se construye una solución mediante la aplicación de un procedimiento voraz aleatorizado.
Se crea iterativamente una lista restringida de candidatos formada por los candidatos más prometedores para ser incorporados a la solución parcial inicialmente vacía.
La lista restringida de candidatos se denota por $LRC$ y, en cada iteración, se selecciona de ella al azar uno de sus elementos que se incorpora a la solución parcial~\cite{Grasp2}.
Para el $k$LSFP utilizamos de una lista restringida de candidatos basada en la evaluación de las etiquetas que se incorporan a $ LRC_\alpha$~\cite{MLSTP-EJOR}.
Es una extensión del criterio clásico aplicado en GRASP, que consiste en colocar en la lista sólo las etiquetas candidatas que tienen una valoración
(el número de componentes conexas en el caso del $k$LSFP) no mayor que un umbral definido por el usuario~\cite{Grasp2}.

En nuestra implementación, se aplica una aleatorización completa para elegir la etiqueta inicial a incorporar la solución vacía.
Esto corresponde a establecer en la primera iteración el umbral en $+\infty$, lo que significa que la lista de candidatos está formada por todas las etiquetas del modelo.
Para las restantes iteraciones, para cada nueva etiqueta a añadir, la lista $LRC$ se forma considerando sólo las etiquetas que dan el número mínimo de componentes conexas en la etapa específica,
con el fin de intensificar al máximo el proceso de búsqueda.
Esto significa que se fija el umbral al número mínimo de componentes conexas obtenidas por las posibles etiquetas a añadir ~\cite{MLSTP-EJOR}.
Es decir, las etiquetas que dan lugar al menor número de componentes conexas en cada paso constituyen la lista restringida de candidatos $LRC$.
Luego, en cada iteración se selecciona aleatoriamente una etiqueta de la lista restringida de candidatos $LRC$ que se añade a la solución actual.
Se actualiza la lista restringida de candidatos para la nueva solución parcial y se itera el proceso.
La fase constructiva finaliza cuando se obtenga una solución con $ {k}$ etiquetas, o que dé lugar a un subgrafo con solo una componente conexa.

La consiguiente mejora por búsqueda local quita con criterio \emph{greedy} una etiqueta y añade otra etiqueta utilizando el algoritmo MVCA,
de forma que el número total de componentes conexas se reduzca al mínimo.
Esto produce una mayor intensificación del algoritmo.

Este proceso de dos fases es iterativo, continuando hasta que se satisfaga la condición de parada especificada por el usuario.
Los criterios de parada usuales hacen referencia a que se alcance un tiempo de CPU máximo permitido, a un número máximo de iteraciones, o a un número máximo de iteraciones entre dos mejoras sucesivas.
El resultado final aportado por el procedimiento GRASP es la mejor solución encontrada hasta entonces en todas las iteraciones.

\section{La Búsqueda por Entornos Variables} \label{sec-vns}

La idea fundamental en una Búsqueda por Entorno Variable (VNS, {\sl Variable Neighbourhood Search}) es el cambio sistemático en la estructura de entornos para escapar de los mínimos locales~\cite{Hansen1}.
La idea de la Búsqueda por Entorno Variable Basica (BVNS, {\sl Basic Variable Neighbourhood Search}) es explorar entornos cada vez más alejados para iniciar la búsqueda local y así tratar de mejorar el óptimo local alcanzado ~\cite{Hansenetal2010}.
Nuestra implementación de BVNS para el $k$LSFP está inspirada en el éxito de la VNS propuesta para el problema de árbol generador de etiquetado mínimo MLST (\emph{Minimum Labeling Spanning Tree}) ~\cite{MLSTP-EJOR}.
Dado el grafo no dirigido $G=(V,A)$ con $|V|=n$ vértices y $|A|=m$ aristas que están etiquetadas con $|L|=\ell$ etiquetas y un entero positivo ${k}$ como límite para el número de etiquetas a utilizar,
cada solución $C \subseteq L$ se codifica por una cadena binaria (\emph{String}) de la forma $ C = [l_1, l_2, ..., l_ {\ ell}] $, donde $ l_i = 1 $ si la etiqueta $ i $ se incluye en la solución $C$,
y $ l_i = 0 $ en caso contrario.
Se selecciona el conjunto de $ q_{max} $ estructuras de entornos $N_q$, con $ q = 1, 2, ..., q_{max}$, cada una de las cuales es una aplicación del espacio de soluciones en las partes o subconjuntos de soluciones.
El parámetro $ q_{max} $ es el número máximo de estructuras de entornos a utilizar.

La elección más común y más simple de estruturas de entornos para esta metaheurística es una familia de entornos de tamaño creciente;
es decir, tales que $|N_1(C)| <...< |N_q(C)| < ... < |N_{q_{max}}(C)|$, $\forall C \subseteq L$.
Sea $ \rho(C_1, C_2) = | C_1 \Delta C_2 | = \ sum_ {i = 1}^{\ ell} \lambda_i $ la distancia de Hamming entre dos soluciones $C_1$ y $C_2$,
donde $ \lambda_i = 1 $ si la etiqueta $ i $ se incluye en una de las dos soluciones, pero no en la otra, y $\lambda_i = 0 $ en caso contrario, $ \forall i = 1, 2, ..., \ell $.
El conjunto de soluciones del $q$-ésimo entorno de una solución de $C$ se establece por $ N_q(C) = \{ C' \subseteq L: \rho(C', C)=q\}$, $\forall q = 1,...,q_{max}$.

La VNS básica que hemos implementado comienza por una solución inicial $C^*$.
Este conjunto inicial de etiquetas se obtiene mediante la inclusión iterativa de etiquetas seleccionadas de forma aleatoria en un conjunto inicialmente vacío hasta tener $ {k} $ etiquetas.
A continuación, el algoritmo aplica la fase de \emph{agitación} desde $C^*$, que consiste en la selección aleatoria de una solución $C'$ del entorno $ N_q(C^*) $ de la solución actual $C^* $,
dejando variar al parámetro $ q $ desde $ 1 $ hasta $ q_{max} $ a lo largo de la ejecución.
Para construir una solución $ C' \in N_q (C^*) $, el proceso de agitación comienza con $C' = C^*$.
En primer lugar, se quitan $q$ etiquetas al azar de $C'$, y, si después de que las etiquetas hayan sido retiradas,
se requiere una mayor expansión de la estructura de entorno (es decir, en el caso de que $ q > |C'| $)
se incluyen nuevas etiquetas adicionales en $C'$ elegidas al azar del conjunto de etiquetas sin usar $ (L-C') $, ~\cite{MLSTP-EJOR}.
El pseudocódigo del proceso de agitación se describe en el Algoritmo \ref{Agitacion}.

\begin{algorithm}[h!] 
Procedimiento \textbf{Agitación($N_q(C)$):}\\
$\;$ \\
\Begin{
Sea $C' \leftarrow C$\; $i \leftarrow 1$\;
\While{$i \le q$}{
    \uIf{$i \leq |C|$}{
        Seleccionar al azar una etiqueta $c'\in C'$\;
        Borrar la etiqueta $c'$ del conjunto de etiquetas usadas: $C'\leftarrow C' - \{c'\}$ \;
        }
    \Else{
        Seleccionar al azar una etiqueta no usada $c'$, i.e. $c'\in (L-C)$\;
        Añadir la etiqueta $c'$ al conjunto de etiquetas usadas: $C'\leftarrow C'\cup\{c'\}$\;
        }
    $i \leftarrow i + 1$ \;
    }
    Actualizar $H'=(V,A(C'))$ y $Comp(C')$\;}
\caption{Agitación de la BVNS para el $k$LSFP.} \label{Agitacion}
\end{algorithm}

Después de la agitación se aplica la búsqueda local con la solución $C'$ como solución inicial.
Esta búsqueda intenta eliminar cada una de las etiquetas de $C'$, y luego agregar nuevas etiquetas siguiendo el método voraz MVCA,
para reducir al mínimo el número de componentes conexas, hasta que se alcance $k$ etiquetas.
Con esta búsqueda local, se obtiene un nuevo conjunto $C''$ de etiquetas.
Si con $C''$ no se ha obtenido una mejora (es decir, si $ | Comp(C'') | \geq | Comp(C^*) | $, la amplitud de la agitación se incrementa ($ q \leftarrow q + 1 $),
dando lugar a una mayor diversificación del proceso de búsqueda.
El pseudocódigo de la búsqueda local se describe en el Algoritmo \ref{LS}.

\begin{algorithm}[h!] 
Procedimiento \textbf{Búsqueda-local$(C')$:}\\
$\;$ \\
\Begin{
Sea $C'' \leftarrow C'$\; $i \leftarrow 1$\;
\While{$i \le |C'|$}{
    Eliminar la etiqueta $i$-ésima de $C'$; sea $C''$: $C'' \leftarrow C'-\{c_i\}$\;
    \While{$(Comp(C'')>1)$ \& $(|C|<k)$}{
        Seleccionar al azar una etiqueta no usada $c\in (L-C'')$ que minimice $Comp(C'' \cup \{c\})$\;
        Añadir la etiqueta $c$ al conjunto de etiquetas usadas: $C''\leftarrow C'' \cup \{c\}$\;
        Actualizar $H''=(V,A(C''))$ y $Comp(C'')$\;
        }
    $i \leftarrow i + 1 $  \;
    }
    }
\caption{Búsqueda local de la BVNS para el $k$LSFP.} \label{LS}
\end{algorithm}

La VNS básica (BVNS \cite{Hansenetal2010} comienza con una solución inicial $C^*$.
Se aplica el procedimiento de \emph{agitación} descrito en el Algoritmo \ref{Agitacion} desde $C^*$ para obtener $C' \in N_q(C^*)$
y desde $C'$ como solución inicial se ejecuta la búsqueda local descrita en Algoritmo \ref{LS} para obtener $C''$.
Si $C''$ mejora a $C^*$ (es decir, si $ | Comp(C'')| < |Comp(C^*)| $) la sustituye como solución actual y se reinicia la amplitud $q$ de la agitación a 1; ($ q \leftarrow 1 $).
En otro caso, se incrementa en una unidad la amplitud de la agitación; $q \leftarrow q + 1$.
De esta forma la amplitud de la agitación $ q $ varía desde $ 1 $ hasta un valor máximo $ q_{max} $ que es el parámetro clave de la BVNS.
El proceso de cambiar la estructura de entornos cuando la búsqueda local alcanza un mínimo local que no mejora la mejor solución hasta entonces
representa la idea central de la VNS básica al proporcionar una diversificación creciente.
Este procedimiento se repite iterativamente hasta que se alcancen las condiciones de parada predeterminadas, proporcionando al final la mejor solución $C^*$ como salida.
El pseudocódigo de la BVNS se describe en el Algoritmo 3. 

\begin{algorithm}[h!]
\begin{minipage}[t]{0.39\textwidth}
\caption{BVNS para el $k$LSF}
\KwIn{Un grafo no dirigido $G=(V,A)$ con $|V|=n$ vértices y $|A|=m$ aristas que están etiquetadas con $|L|=\ell$ etiquetas y un entero positivo ${k}$ ;}
\KwOut{Un bosque $F$;}
$\;$ \\
\Begin{
 $C^* \leftarrow generar-al-azar()$\;
    \Repeat{condiciones de parada}
    {
       tomar $q\leftarrow 1$ y $q_{max} \leftarrow |C^*|$\;
       \While{$q < q_{max}$}
       {
        $C' \leftarrow$  agitación($N_q(C^*)$)\;
        $C'' \leftarrow$ búsqueda-local($C'$)\;
          \uIf{$|C''|<|C^*|$}
          {
              moverse: $C^*\leftarrow C''$\;
              reinicializar la amplitud de agitación $q \leftarrow  1$ \;
              actualizar la máxima amplitud de agitación $q_{max} \leftarrow  |C^*|$\;}
          \Else
          {
               aumentar la amplitud de agitación: $q\leftarrow q+1$\;
               }
          }
    }
   Tomar un bosque generador $F$ del grafo $G(C)=(V,A(C))\;$
   }
   \end{minipage}
   \label{BVNS}
\end{algorithm}

\subsection{Ajuste de la amplitud de agitación}

El parámetro $q_{max}$ determina la amplitud que llega a alcanzar el proceso de agitación en la VNS básica.
El valor de este parámetro es relevante para modular la VNS básica y así conseguir un balance adecuado entre la potencia de intensificación y la capacidad de diversificación ~\cite{Hansen1}.
La selección de un valor pequeño para $ q_{max} $ produce una alta potencia intensificadora y una baja capacidad diversificadora, dando lugar a un algoritmo rápido,
pero con una alta probabilidad de quedar atrapado en un mínimo local no óptimo.
Por el contrario, un valor grande de $ q_{max} $ disminuye la capacidad de intensificación y aumenta la capacidad de la diversificación, lo que da lugar a un algoritmo más lento,
pero con mayor facilidad para escapar de los mínimos locales subóptimos.
Con objeto de estudiar la relevancia del parámetro y optar por la mejor alternativa se establecieron tres tipos de estrategia.

\begin{enumerate}

\item La primera estrategia consiste en elegir el parámetro $q_{max}$ fijo para todas las ejecuciones del algoritmo ($q_{max} = \alpha$), independientemente de la instancia a la que se aplica. Por tanto, la potencia exploradora de la agitación es fija.

\item La segunda estrategia consiste en elegir el parámetro $q_{max}$ dependiendo de la dificultad o magnitud de la instancia. En este caso optamos por tomar la cota superior $k$ del número de etiquetas a usar como una medida de la que depende directamente el tamaño del espacio de soluciones. Por tanto aplicamos una fórmula del tipo $q_{max} = \alpha \cdot k$. La potencia exploradora de la agitación se hace depender del tamaño del espacio de soluciones.

\item La tercera estrategia consiste en elegir el parámetro $q_{max}$ dependiendo del estado en que se encuentre la búsqueda. Como reflejo del estado en que se encuentra la búsqueda elegimos el tamaño de la solución actual. La fórmula para establecer el tamaño de la agitación es $q_{max} = \alpha \cdot |C|$, siendo $C$ la solución actual. Por tanto, la potencia exploradora de la agitación se hace depender del estado en el que se encuentra el proceso de búsqueda medido en el valor de la función objetivo.

\end{enumerate}

Para evaluar el rendimiento de las estrategias descritas para la BVNS se utilizó un subconjunto de las instancias propuestas en la literatura para este tipo de problemas (ver ~\cite{Cerulli1}).
Se trata de conjuntos de instancias generadas aleatoriamente a partir del número de nodos ($n$) y el número de etiquetas ($\ell$).
Se obtuvieron conjuntos de instancias para diferentes combinaciones de los parámetros $n$, $d$ y $\ell$. El número $n$ de vertices varia de 100 a 500 y el número $\ell$ de etiquetas de $0,25n$ a $1.25n$.
Para establecer el número de aristas se fijó la densidad del grafo en $d = 0.5$ por lo que el número máximo de aristas de cada instancia es  $ m = 0.5 n(n-1)$.
El valor del parámetro $ k $ para adaptar estas instancias al $k$LSFP se determinó experimentalmente como $ \lfloor n/2^j \rfloor $, donde $ j $ es el valor más pequeño de tal manera que las instancias generadas no formaran una única solución conexa una vez resuelto el MVCA ~\cite{CerulliKLSF}.
Para cada combinación de $ n $, $d$, $ \ell$ y $k$ se generaron al azar 10 instancias distintas.
La cota $k$ para el número de etiquetas es un factor de importancia crucial para crear instancias significativas ya que, si se elige demasiado grande (en particular, mayor que el número de etiquetas necesarias para obtener un MLST), podría ser fácil para la heurística para detectar una sola componente conexa como solución óptima en un tiempo extremadamente corto. Por otro lado, si el valor de $k$ es demasiado pequeño, el problema podría ser igualmente fácil, ya que el espacio de búsqueda podría ser recorrido muy rápidamente.

Para analizar las etrategias diferentes de selección de la máxima amplitud de la agitación $q_{max}$ descritas y optar por la que tenga mejor rendimiento se realizaron experimentos con 8 de estos conjuntos de instancias diferentes para cada una de las combinaciones de los números de vértices $n = 100$ y $n = 200$ y números de etiquetas $ \ell = 0.25 n$, $\ell = 0.5 n $, $\ell = n $, y $\ell = 1.25n $ para un total de $ 80 $ casos.
En todos estos casos se ejecutaron los algoritmos por un máximo de 60 segundos de tiempo CPU y el tiempo que aperece en las tablas corresponde al tiempo que emplearon en encontrar la mejor solución expresado en milisegundos.
Las instancias están disponibles en \texttt{http://www.sergioconsoli.com/MLSTP.htm}.
Todos los cálculos se hicieron en un Mac OSX 10.9 con microprocesador a 3,0 GHz y con 8 GB de RAM.

Se compararon 5 casos particulares de las tres estrategias anteriores eligiendo 5 valores para el parámetro $\alpha$ en cada una de ellas. Así para la estrategia 1 se consideraron los valores de $\alpha$ en $\{5, 10, 15, 20, 25\}$. Para la estrategia 2 se consideraron los valores de $\alpha$ en $\{0.1, 0.3, 0.5, 0.7, 0.9\}$. Finalmente, para la estrategia 3 se consideraron los valores de $\alpha$ en $\{1/3, 2/3, 3/3, 4/3, 5/3\}$.
Las 15 estrategias de actualización del parámetro $q_{max}$ correspondientes que dan lugar a 15 versiones de la BVNS se resumen en la Tabla \ref{Estrategias}.
Los resultados computacionales obtenidos con estas 15 versiones de la BVNS se muestran en las Tablas \ref{Estrategia1} a \ref{Estrategia3}.

\begin{table}[ht]
\begin{tabular}{||c|c||c|c|c|c|c||}
\hline\hline
Num. & $q_{max}$  & \multicolumn{5
}{|c||}{Valores de $\alpha$}\\
\hline
1 & $= \alpha$ & 5 & 10 & 15 & 20 & 25\\
2 & $= \alpha\cdot k$ & 0.1 & 0.3 & 0.5 & 0.7 & 0.9\\
3 & $= \alpha\cdot |C|$ & 1/3 & 2/3 & 3/3 & 4/3 & 5/3\\
\hline\hline
\end{tabular}
\caption{Estrategias de selección de la amplitud de agitación}
\label{Estrategias}
\end{table}

Los resultados computacionales obtenidos con cada una de las tres estrategia propuestas se pueden ver, respefctivamente, en la Tabla~\ref{Estrategia1}, en la Tabla~\ref{Estrategia2} y en Tabla~\ref{Estrategia3}

\begin{table*}[ht]
\begin{tabular}{||c|c|c| |c|c| |c|c| |c|c| |c|c| |c|c||}
\hline
\multicolumn{3}{||c||}{Parámetros}&\multicolumn{2}{|c||}{VNS-01-5}&\multicolumn{2}{|c||}{VNS-01-10}&\multicolumn{2}{|c||}{VNS-01-15}&\multicolumn{2}{|c||}{VNS-01-20}&\multicolumn{2}{|c||}{VNS-01-25} \\
\hline
$n$	&$\ell$	&	$k$	&	Obj	&	Tiempo	&	Obj	&	Tiempo	&	Obj	&	Tiempo	&	Obj	&	Tiempo	&	Obj	&	Tiempo	\\
100	&	25	&	3	&	2.1	&	20.5	&	2.1	&	1.2	    &	2.1	&	0.7	    &	2.1	&	0.9	    &	2.1	&	1.9	\\
100	&	50	&	3	&	5.2	&	7.8	    &	5.2	&	4.6	    &	5.2	&	4.5	    &	5.2	&	4.4	    &	5.2	&	5.7	\\
100	&	100	&	4	&	8.3	&	49.8	&	8.3	&	24.6	&	8.3	&	164.7	&	8.3	&	19.9	&	8.3	&	85.8	\\
100	&	125	&	5	&	4.1	&	106.4	&	4.1	&	196.7	&	4.1	&	220.5	&	4.1	&	68.1	&	4.1	&	195.7	\\
200	&	50	&	3	&	2.0	&	128.6	&	2.0	&	37.9	&	2.0	&	109.6	&	2.0	&	98.9	&	2.0	&	45.7	\\
200	&	100	&	4	&	2.2	&	741.6	&	2.2	&	942.7	&	2.2	&	1085.4	&	2.2	&	797.5	&	2.2	&	933.2	\\
200	&	200	&	5	&	7.0	&	5379.6	&	6.8	&	4923.4	&	6.8	&	2072.8	&	6.8	&	4700.5	&	6.8	&	3580.9	\\
200	&	250	&	6	&	4.4	&	1218.3	&	3.9	&	9701.9	&	4.2	&	12522.3	&	3.9	&	10862.4	&	4.0	&	14854.4	\\
\hline
\end{tabular}
\caption{Resultados para la estrategia 1}
\label{Estrategia1}
\end{table*}

\begin{table*}[ht]
\begin{tabular}{||c|c|c| |c|c| |c|c| |c|c| |c|c| |c|c||}
\hline
\multicolumn{3}{||c||}{Parámetros}&\multicolumn{2}{|c||}{VNS-02-1}&\multicolumn{2}{|c||}{VNS-02-3}&\multicolumn{2}{|c||}{VNS-02-5}&\multicolumn{2}{|c||}{VNS-02-7}&\multicolumn{2}{|c||}{VNS-02-9} \\
\hline
$n$	&$\ell$	&	$k$	&	Obj	&	Tiempo	&	Obj	&	Tiempo	&	Obj	&	Tiempo	&	Obj	&	Tiempo	&	Obj	&	Tiempo	\\
100	&	25	&	3	&	2.1	&	1.5	&	2.1	&	1.4	&	2.1	&	1.0	&	2.1	&	1.1	&	2.1	&	1.6	\\
100	&	50	&	3	&	5.2	&	30.2	&	5.2	&	4.1	&	5.2	&	4.4	&	5.2	&	62.3	&	5.2	&	4.7	\\
100	&	100	&	4	&	8.3	&	68.9	&	8.3	&	125.9	&	8.3	&	44.1	&	8.3	&	73.8	&	8.3	&	83.2	\\
100	&	125	&	5	&	4.1	&	42.4	&	4.1	&	68.3	&	4.1	&	155.9	&	4.1	&	219.1	&	4.1	&	152.7	\\
200	&	50	&	3	&	2.0	&	118.9	&	2.0	&	69.7	&	2.0	&	38.2	&	2.0	&	127.9	&	2.0	&	17.4	\\
200	&	100	&	4	&	2.2	&	1943.9	&	2.2	&	1305.3	&	2.2	&	1407.6	&	2.2	&	1269.1	&	2.2	&	1068.9	\\
200	&	200	&	5	&	6.8	&	8298.7	&	6.8	&	3231.1	&	6.8	&	2156.3	&	6.8	&	3027.8	&	6.8	&	6570.6	\\
200	&	250	&	6	&	4.1	&	8638.9	&	4.1	&	11113.5	&	4.1	&	11183.2	&	4.0	&	17258.8	&	4.0	&	13368.5	\\
\hline
\end{tabular}
\caption{Resultados para la estrategia 2}
\label{Estrategia2}
\end{table*}

\begin{table*}[ht]
\begin{tabular}{||c|c|c| |c|c| |c|c| |c|c| |c|c| |c|c||}
\hline
\multicolumn{3}{||c||}{Parámetros}&\multicolumn{2}{|c||}{VNS-03-1/3}&\multicolumn{2}{|c||}{VNS-03-2/3}&\multicolumn{2}{|c||}{VNS-03-3/3}&\multicolumn{2}{|c||}{VNS-03-4/3}&\multicolumn{2}{|c||}{VNS-03-5/3} \\
\hline
$n$	&$\ell$	&	$k$	&	Obj	&	Tiempo	&	Obj	&	Tiempo	&	Obj	&	Tiempo	&	Obj	&	Tiempo	&	Obj	&	Tiempo	\\
100	&	25	&	3	&	7.5	&	0.6	&	7.5	&	0.7	&	2.1	&	3.1	&	2.1	&	1.0	&	2.3	&	1.7	\\	
100	&	50	&	3	&	15.4	&	0.6	&	15.4	&	0.8	&	5.3	&	2.6	&	5.2	&	4.2	&	5.8	&	4.5	\\	
100	&	100	&	4	&	29.3	&	0.4	&	8.8	&	40.1	&	8.7	&	29.9	&	8.3	&	11.9	&	8.3	&	44.7	\\	
100	&	125	&	5	&	29.2	&	0.7	&	4.9	&	572.4	&	4.5	&	23.9	&	4.1	&	36.1	&	4.1	&	38.8	\\	
200	&	50	&	3	&	9.0	&	0.7	&	9.0	&	0.7	&	2.1	&	64.1	&	2.0	&	10.2	&	2.3	&	19.7	\\	
200	&	100	&	4	&	14.3	&	0.2	&	3.5	&	362.4	&	2.6	&	658.2	&	2.2	&	731.1	&	2.2	&	2199.1	\\	
200	&	200	&	5	&	33.7	&	0.5	&	9.1	&	238.1	&	7.2	&	673.9	&	6.8	&	1519.1	&	6.8	&	8636.5	\\	
200	&	250	&	6	&	36.2	&	0.9	&	6.3	&	357.2	&	4.6	&	7979.3	&	3.9	&	2789.2	&	4.1	&	4600.1	\\	
\hline
\end{tabular}
\caption{Resultados para la estrategia 3}
\label{Estrategia3}
\end{table*}

Los resultados obtenidos con la estrategia 1 son bastante pobres aunque los valores del objetivo para los tamaños 20 y 25 son aceptables, aunque no el tiempo de ejecución.
Los resultados Los resultados computacionales obtenidos con la estrategia 2 ya pasan a ser mejores pero se detecta una excesiva diversificación para valores grandes de $\alpha$ y los tiempos de ejecución son muy altos.
Para la estrategia 3 se observa un doble comportamiento. Cuando $\alpha$ es bajo, el algoritmo termina de manera prematura con soluciones de baja calidad, adoleciendo de capacidad de exploración.
Cuando $\alpha$ es mayor que uno se observan mejores resultados porque la agitación se amplía.
Sin embargo cuando $\alpha$ crece demasiado se produce un pico de diversificación dando lugar a malos resultados que se traducen en un tiempo excesivo sin aportar soluciones de muy alta calidad.
Como consecuencia de ello, tenemos que la tercera estrategia con una agitación de amplitud dependiente del tamaño de la solución actual con una constante de proporcionalidad directa de 4/3,
$ q_ {max} = 4|C|/3 $, produce los mejores resultados. Esta conclusión coincide con lo ya observado en ~\cite{MLSTP-EJOR} donde se muestra que el parámetro $ q_ {max} = (4| C | / 3) $
da un buen balance entre exploración y explotación para el MLSTP.

\section{Comparación entre metaheurísticas} \label{resultados}

Para evaluar el rendimiento y eficiencia del algoritmo propuesto en comparación con los descritos en la Sección \ref{otrosalg} se usaron también las instancias propuestas en \cite{CerulliKLSF}.
Se realizaron experimentos con 20 conjuntos de instancias diferentes con un número de vértices $ n \in \{ 100, 200, 300, 400, 500 \} $, y número de etiquetas $ \ell \in \{ 0,25n, 0.5n, n, 1.25n \}$.
Para cada combinación de $ n$ y $ \ell $, se dispone de 10 instancias distintas, para un total de $ 200 $ casos.
Estas instancias también están disponibles en \texttt{http://www.sergioconsoli.com/MLSTP.htm}.

Los resultados obtenidos se ofrecen en la Tabla~\ref{GroupRes}. Las tres primeras columnas muestran los parámetros que caracterizan los diferentes conjuntos de instancias ($ n $, $ \ ell $ y $k$),
mientras que las columnas restantes dan los resultados obtenidos con los algoritmos descritos.
Para cada conjunto de instancias, la calidad de la solución obtenida (Obj) se evalúa como el número medio de componentes conexas entre las 10 instancias del problema.
Un tiempo de CPU máximo permitido de 60 o 600 segundos fue elegido como condición de parada de todas los metaheurísticas,
determinado experimentalmente con respecto a la dimensión de los conjuntos de datos considerados
(es decir,  1 minuto para $ n \leq 200 $, y 10 minutos para $ n > 200 $).
Para el algoritmo genético fijamos el número de generaciones para cada instancia de tal manera que los cómputos lleven aproximadamente al máximo tiempo de CPU permitido.
Todas las metaheurísticas se ejecutan hasta el tiempo máximo de CPU que se alcanza y, en cada caso, se registra la mejor solución.
Los tiempos computacionales de la Tabla~\ref{GroupRes} de las columnas etiquetadas con ``Tiempo" son los tiempos medios en los que se obtienen las mejores soluciones.
En el caso de que no se haya obtenido ninguna solución en 3 horas por el método exacto se denota por NF en la Tabla~\ref{GroupRes}.
El rendimiento de un algoritmo se puede considerar mejor que otro si obtiene menor valor objetivo promedio o con un valor objetivo promedio similar pero con un menor tiempo computacional.

\begin{table*}[ht]
\begin{tabular}{|c|c|c||cc |cc |cc |cc |cc|}
\hline
\multicolumn{3}{|c||}{$d = 0.5$}&\multicolumn{2}{c|}{\bf{EXACTO}}&\multicolumn{2}{c|}{\bf{PM}}&\multicolumn{2}{c|}{\bf{GA}}&\multicolumn{2}{c|}{\bf{GRASP}}&\multicolumn{2}{c|}{\bf{BVNS}}  \\
$n$ & $\ell$ & ${k}$ & Obj & Tiempo & Obj & Tiempo & Obj & Tiempo & Obj & Tiempo & Obj & Tiempo \\
\hline
\hline
 100    & 25   & 3    &     2,1		&	59,4	&	2,1		&	6,8		&	2,1		&	47,2	&	2,1		&	7,7		&	2,1		&	1,1		\\
 100    & 50   & 3    &     5,2		&	84,3	&	5,2		&	35,7	&	5,2		&	62,6	&	5,2		&	19,4	&	5,2		&	4,2		\\
 100    & 100  & 4    &     8,3		&	1100 	&	8,3		&	125,7	&	8,3		&	215,2	&	8,3		&	108,5	&	8,3		&	16,3	\\
 100    & 125  & 5    &     4,1		&	44400 	&	4,3		&	217,2	&	4,1		&	281,7	&	4,3		&	143,6	&	4,1		&	29,5	\\
\cline{1-13}                    	                                      	
 200    & 50   & 3    &     2	  	&	481,4	&	2	  	&	65,1	&	2		&	458,9	&	2		&	30,3	&	2		&	8,4		\\
 200    & 100  & 4    &     2,2		&	2500 	&	2,4		&	699,3	&	2,7		&	1100	&	2,4		&	319,1	&	2,2		&	243,6	\\
 200    & 200  & 5    &     6,6		&	978900	&	6,9		&	5300	&	7		&	11300	&	7		&	775,2	&	6,6		&	1900	\\
 200    & 250  & 6    &     NF		&	NF		&	4,3		&	9200	&	4,4		&	12700	&	4,3		&	10400	&	3,9		&	7500	\\
\cline{1-13}                    	                                      	
 300    & 75   & 3    &     2	  	&	793,8	&	2		&	295,5	&	2		&	1600	&	2		&	132,3	&	2		&	60,2	\\
 300    & 150  & 4    &     3,6		&	16700	&	3,8		&	2600	&	4		&	9100	&	3,9		&	10100	&	3,6		&	4400	\\
 300    & 300  & 5    &     NF		&	NF		&	13,5	&	12200	&	13,6	&	14100	&	13,5	&	13100	&	12,9	&	9700	\\
 300    & 375  & 6    &     NF		&	NF		&	10,3	&	18900	&	10,7	&	16500	&	10,1	&	5900	&	9,7		&	41100	\\
\cline{1-13}                    	                                      	
 400    & 100  & 3    &     2,3		&	1500	&	2,3		&	955,2	&	2,4		&	7200	&	2,3		&	365,1	&	2,3		&	236,9	\\
 400    & 200  & 3    &     31,9	&	23300	&	31,9	&	6500	&	31,9	&	10300	&	31,9	&	3600	&	31,9	&	503,8	\\
 400    & 400  & 5    &     NF		&	NF		&	21,1	&	38300	&	21,6	&	42900	&	21,5	&	7900	&	20,9	&	175000	\\
 400    & 500  & 6    &     NF		&	NF		&	18,8	&	58800	&	18,7	&	26900	&	18,3	&	33300	&	17,7	&	39500	\\
\cline{1-13}                    	                                      	
 500    & 125  & 3    &     2,9		&	21200	&	2,9		&	2200	&	3,1		&	22500	&	2,9		&	693,8	&	2,9		&	617,1	\\
 500    & 250  & 4    &     5,8		&	193900	&	6,1		&	19700	&	6,3		&	36800	&	6,1		&	9300	&	5,8		&	25200	\\
 500    & 500  & 6    &     NF		&	NF		&	8,8		&	95700	&	9,1		&	117800	&	8,8		&	50300	&	8,1		&	166400	\\
 500    & 625  & 7    &     NF		&	NF		&	7,7		&	155100	&	8,9		&	136900	&	7,7		&	140600	&	7		&	91700	\\
\hline
\hline
\multicolumn{3}{|r||}{TOTAL:}& -   &-  &164,7       &426900     &168,1    &468800    &164,6 &287100 & 159,2 &406600  \\
\hline
\end{tabular}
\caption{Resultados para $|V|=100, 200, 300, 400, 500$ y $|L|=0.25|V|$, $0.5|V|$, $|V|$, $1.25|V|$}
\label{GroupRes}
\end{table*}

En la Tabla~\ref{GroupRes} se observa que la BVNS es capaz de obtener muy alto rendimiento para el problema $k$LSF. En todos los conjuntos de datos se obtiene muy rápidamente las soluciones con la mejor calidad.
Con respecto al tiempo de ejecución, unas veces fue más lento que las otras heurísticas pero debido a que obtiene mucho mejores soluciones de calidad (por ejemplo, ver las instancias de
[$ n = 200, \ell = 200 $], [$ n = 300, \ell = 375 $], [$ n = 400, \ell = 500 $], [$ n = 500, \ell = 500 $]).
Cuando la calidad media de la solución obtenida con la BVNS fue comparable a la obtenida por los otros procedimientos, empleó un tiempo computacional menor.
En cuanto a las otras heurísticas, el algoritmo genético no tuvo muy buen rendimiento tanto con respecto a la calidad de la solución y como con respecto al tiempo de cómputo, lo que confirma, en general, un mal comportamiento de las metaheurísticas poblacionales al abordar a esta clase de problemas (véase, por ejemplo~\cite{CerulliKLSF,MLSTP-EJOR,MLSteiner-ANOR}).
El Método Piloto registró un mejor comportamiento que el algoritmo genético con respecto a la calidad media de las soluciones, pero los tiempos de ejecución eran aún muy grandes, lo que indica una dispersión excesiva con respecto a la diversificación. GRASP parece ser una buena opción en cuanto al tiempo de ejecución, pero la calidad media de las soluciones no fue tan buena como con la BVNS. Es interesante notar también que en todos los casos de problemas para los que el método exacto obtuvo la solución óptima, sólo la BVNS también produjo siempre la misma solución, y con un tiempo de cómputo muy corto. La BVNS proporciona el mejor compromiso entre las capacidades de intensificación y diversificación para el problema considerado.
Resumiendo los resultados demuestran que la BVNS es una metaheurística muy eficaz para el problema $k$LSF, produciendo soluciones de alta calidad en tiempos computacionales de ejecución cortos.

\section{Conclusiones} \label{sec-conclusions}

En este trabajo hemos considerado el problema del bosque generador $k$-etiquetado ($k$LSFP),
un problema $NP$-duro que resulta de una extensión del problema del árbol generador de etiquetado mínimo (MLSTP) al caso en el que
se impone una cota superior ${k}$ al número de etiquetas a utilizar; por lo que el propósito es obtener la solución con el mínimo número de componentes conexas que no violen esta restricción.
Ya que el problema $k$LSF es $NP$-duro, son interesantes las heurísticas y metaheurística que permitan abordar con garantías de rendimiento.
Se describe una BVNS para la que se ajusta el parámetro clave que fija la máxima amplitud que puede alcanzar el proceso de agitación.
Se comprueba experimentalmente que la mejor opción es hacer depender dicho parámetro del estado del proceso de búsqueda, conclusión que podría extenderse a otros problemas.
Al menos se presume que una fórmula de actualización similar sería aplicable a los problemas en los que el tamaño de la solución refleja el estado del proceso de búsqueda.
Con este ajuste del parámetro, la BVNS se comparó experimentalmente sobre una amplia gama de instancias del $k$LSFP, con otras metaheurísticas propuestas en la literatura para el problema: el Método Piloto, un Algoritmo Genético, y un procedimiento de búsqueda voraz aleatoria adaptativa (GRASP), junto con un enfoque exacto de resolución.
Nuestros experimentos computacionales muestran que la BVNS debidamente ajustada supera a todas las otras metaheurísticas y es rápido, simple, y particularmente eficaz para el problema $k$LSF.


\section*{Agradecimientos}
Este trabajo está parcialmente financiado por el proyecto  TIN2012-32608 del Ministerio de Economía y Competitividad.

\nocite{*}
\bibliographystyle{maeb2015}

\begin{thebibliography}{10}
\expandafter\ifx\csname url\endcsname\relax
  \def\url#1{\texttt{#1}}\fi
\expandafter\ifx\csname urlprefix\endcsname\relax\def\urlprefix{URL }\fi
\newcommand{\enquote}[1]{``#1''}

\bibitem{bruggemann}
T. Br\"uggemann, J. Monnot, G. J. Woeginger, \emph{Local search for the minimum
label spanning tree problem with bounded colour classes}, Operations Research Letters 31 (2003) 195--201.


\bibitem{CerulliKLSF}
Cerulli, R., A.~Fink, M.~Gentili and A.~Raiconi, \emph{The k-labeled spanning
  forest problem}, Procedia - Social and Behavioral Sciences \textbf{108}
  (2014), pp.~153--163.

\bibitem{Cerulli2}
Cerulli, R., A.~Fink, M.~Gentili and S.~Vo{\ss}, \emph{Extensions of the
  minimum labelling spanning tree problem}, Journal of Telecommunications and
  Information Technology \textbf{4} (2006), pp.~39--45.

\bibitem{Cerulli1}
Cerulli, R., A.~Fink, M.~Gentili and S.~Vo{\ss}, \emph{Metaheuristics comparison for the minimum
labelling spanning tree problem}, in: B. L. Golden, S. Raghavan, E. A. Wasil (Eds.),
\emph{The Next Wave on Computing, Optimization, and Decision Technologies}, Springer-Verlag, New York, 2005, pp. 93--106.


\bibitem{Chang}
Chang, R.~S. and S.~J. Leu, \emph{The minimum labelling spanning trees},
  Information Processing Letters \textbf{63} (1997), pp.~277--282.

\bibitem{Chwatal2010}
A. M. Chwatal, G. R. Raidl, \emph{Solving the minimum label spanning tree
problem by ant colony optimization}, Proceedings of the 7th International
Conference on Genetic and Evolutionary Methods (GEM 2010) (2010) 91--97. Las Vegas, Nevada.

\bibitem{MLSTP-EJOR}
Consoli, S., K.~Darby-Dowman, N.~Mladenovi\'c and J.~A. Moreno-Pérez,
  \emph{Greedy randomized adaptive search and variable neighbourhood search for
  the minimum labelling spanning tree problem}, European Journal of Operational
  Research \textbf{196} (2009), pp.~440--449.

\bibitem{MLSteiner-ANOR}
Consoli, S., K.~Darby-Dowman, N.~Mladenovi\'c and J.~A. Moreno-Pérez,
  \emph{Variable neighbourhood search for the minimum labelling {Steiner} tree
  problem}, Annals of Operations Research \textbf{172} (2009), pp. 71--96.

\bibitem{MLSTP-hybrid}
S. Consoli, J. A. Moreno-P\'erez, \emph{Solving the minimum labelling spanning
tree problem using hybrid local search}, Electronic Notes in Discrete Mathematics  \textbf{39} (2012) pp. 75--82.

\bibitem{GAs}
Goldberg, D.~E., K.~Deb and B.~Korb, \emph{Don't worry, be messy}, in:
  \emph{Proceedings of the 4th International Conference on Genetic Algorithms}
  (1991), pp. 24--30.

\bibitem{Hansen1}
Hansen, P. and N.~Mladenovi\'c, \emph{Variable neighbourhood search}, Computers
  and Operations Research \textbf{24} (1997), pp.~1097--1100.

\bibitem{Hansenetal2010}
Hansen, P., N.~Mladenovi\'c, and J.A.~Moreno-P\'erez, \emph{Variable neighbourhood
search: methods and applications}, Annals of Operations Research \textbf{175} (2010), pp. 367--407.

\bibitem{Krumke}
Krumke, S. O. and H. C. Wirth, \emph{On the minimum label spanning tree problem},
Information Processing Letters \textbf{66} (1998), pp. 81--85.

\bibitem{Miller2006}
Miller, J.~S., \enquote{Multimodal statewide transportation planning: A survey
  of state practices,} Transportation Research Board, National Research
  Council, Virginia, US, 2006.

\bibitem{Grasp2}
Resende, M. G.~C. and C.~C. Ribeiro, \emph{Greedy randomized adaptive search
  procedure}, in: F.~Glover and G.~Kochenberger, editors, \emph{Handbook of
  metaheuristics}, Kluwer Academic Publishers, Norwell, MA, 2003 pp. 219--249.



\bibitem{Golden2}
Xiong, Y., B.~Golden and E.~Wasil, \emph{A one-parameter genetic algorithm for the minimum
labelling spanning tree problem}, IEEE Transactions on Evolutionary Computation \textbf{9}
(2005), pp. 55--60.


\bibitem{Golden3}
Xiong, Y., B.~Golden and E.~Wasil, \emph{Improved heuristics for the minimum
  labelling spanning tree problem}, IEEE Transactions on Evolutionary
  Computation \textbf{10} (2006), pp.~700--703.


\bibitem{Vannes}
Van-Nes, R., \emph{Design of multimodal transport networks: A hierarchical
approach}, Delft University Press, 2002

\end{thebibliography}

\end{document}